\pdfoutput=1 
\RequirePackage{ifpdf}
\documentclass{JINST}
\usepackage[utf8]{inputenc}
\usepackage[english]{babel}
\usepackage{amsfonts}
\usepackage{amssymb}
\usepackage{amsmath}
\usepackage[caption=false]{subfig}
\usepackage{float}
\usepackage{url}
\usepackage{cite}
\usepackage[hidelinks,bookmarks]{hyperref}

\title{\vspace{-15mm}\fontsize{22pt}{10pt}\selectfont\textbf{A dynamic method for charging-up calculations: the case of GEM}} 

\author{P.M.M. Correia$^{a,}$\thanks{Corresponding author.
E-mail: \email{pmcorreia@ua.pt}}~,
C.A.B. Oliveira$^{a,b}$,
C.D.R. Azevedo$^{a}$,
A.L.M. Silva$^{a}$,
R. Veenhof$^c$,
Mythra Varun Nemallapudi$^c$
and J.F.C.A. Veloso$^a$\\
\llap{$^a$}I3N -Physics Department, University of Aveiro,\\
3810-193 Aveiro, Portugal\\
\llap{$^b$}Lawrence Berkeley National Laboratory,\\
One Cyclotron Road, Berkeley, 94720, CA, U.S.A.\\
\llap{$^c$}CERN RD-51 Collaboration, Geneva, Switzerland}

\abstract{\noindent The simulation of \textit{Micro Pattern Gaseous Detectors} (MPGDs) signal response is an important and powerful tool for the design and optimization of such detectors. However, several attempts to simulate exactly the effective charge gain have not been completely successful. Namely, the gain stability over time has not been fully understood.  
\textit{Charging-up} of the insulator surfaces have been pointed as one of the responsible for the difference between experimental and Monte Carlo results. 
This work describes two iterative methods to simulate the \textit{charging-up} in one MPGD device, the Gas Electron Multiplier (GEM). 
The first method uses a constant step-size for avalanches time evolution, very detailed, but slower to compute. 
The second method uses a dynamic step-size that improves the computing time. Good agreement between both methods was achieved. 
Comparison with experimental results shows that \textit{charging-up} plays an important role in detectors operation, explaining the time evolution of the effective gain. However it doesn't seem to be the only responsible for the difference between measurements and Monte Carlo simulations.}


\begin{document}
\selectlanguage{english}
\section*{Introduction}\label{sec:intr}

A considerable amount of work has been done over the last few years to improve the simulations of MPGDs. A full understanding of the micro-physics associated with MPGDs operation is vital for the improvement of such detectors, whose some behaviours are still not completely understood.\cite{COliveira2012JINST,Ozkan2010JINST,spark_simulation}.


Some works report that there is a transient period during which the effective gain changes, after voltages are applied and the detector irradiated\cite{Azmoun200611,THGEM_operation_Ne_CH4}. 
The gain ends up stabilizing after minutes or even hours, depending on the MPGD and the rates of irradiation.

MPGDs were developed to detect radiation, and their main applications are for high energy physics, astrophysics, rare-event searches and medical imaging\cite{mpgd_progression,natal_luz}. 
The \mbox{\textit{Gas Electron Multiplier}} (GEM)\cite{Sauli1997531} has been largely used in many of those applications. 

The device consists of a thin polyimide (insulator) foil typically 50 $\mu$m thick. The foil is covered on both sides with 5 $\mu$m thick layers of a conductor and etched with an hexagonal pattern of holes. The device operates inside a gas medium and suitable electric potentials are applied between the upper and the lower electrodes of the structure. In this way, a very high electric field is created inside the holes. Electrons created in the drift region by interaction of external radiation, travel towards the micro-structure, being focused into the holes and accelerated. They acquire enough energy to ionize atoms/molecules of the gas, creating new ionizations. The secondary electrons undergo the same process while inside the hole and an avalanche ends up being produced, the Townsend avalanche.

The charges produced during multiplication have two possible destinations: they may be collected by conductor electrodes, both those of the GEM itself or of any other readout setup; and a fraction of them ends up accumulating in the insulator surfaces.
The number of simulated avalanches is correlated with the number of primary electrons that undergo to the hole, since we assume that  no charges will drift in the insulator. 

Effective gain is defined as the number of secondary electrons, for each primary electron, that are collected in an electrode plane located below the GEM (figure~\ref{fig:plane_section}).
 
 The electronic affinity of the polyimide usually used in these devices is high, 1.4 eV \cite{poly_affinity}. Once electrons are trapped, it is unlikely that they are able to leave the surface. The same process happens with ions, according to \cite{sessler}. 

The effective gain of the detector strongly depends on the intensity of the electric field produced in the multiplication region. Charges accumulated in the insulator surfaces change locally the electric field, changing the amplification gain. This is known as the charging-up effect in the insulator.

Deposited charges can flow through the insulator surface and insulator bulk under the action of the electric field. 
Previous studies propose that the positive ions are not captured in insulators surfaces, instead they transfer their charge to intrinsic carriers of the insulator, and the conduction should be made by electrons and holes\cite{sessler}. 
The time for charges evacuation is of the order of several hours to days, so we did not include this effect in our method, i.e. all deposited charges will remain in the same surface during the whole simulation time. 
This approach should remain valid if the charging-up process is much faster than the draining of charges.

Simulations of the charging-up influence in the GEM transparency were already reported\cite{Alfonsi20126}. 
In order to study the contribution of the charging-up for the effective gain variations, two methods to simulate the charge accumulation in the detector are presented.
We also compare our results with available experimental data.

\section*{Calculations}\label{sec:gem_thgem}

\subsection*{Geometry}\label{subsec:mpgds_geom}

\begin{figure}[htp]
\centering
\subfloat[]{\includegraphics[width=.45\textwidth]{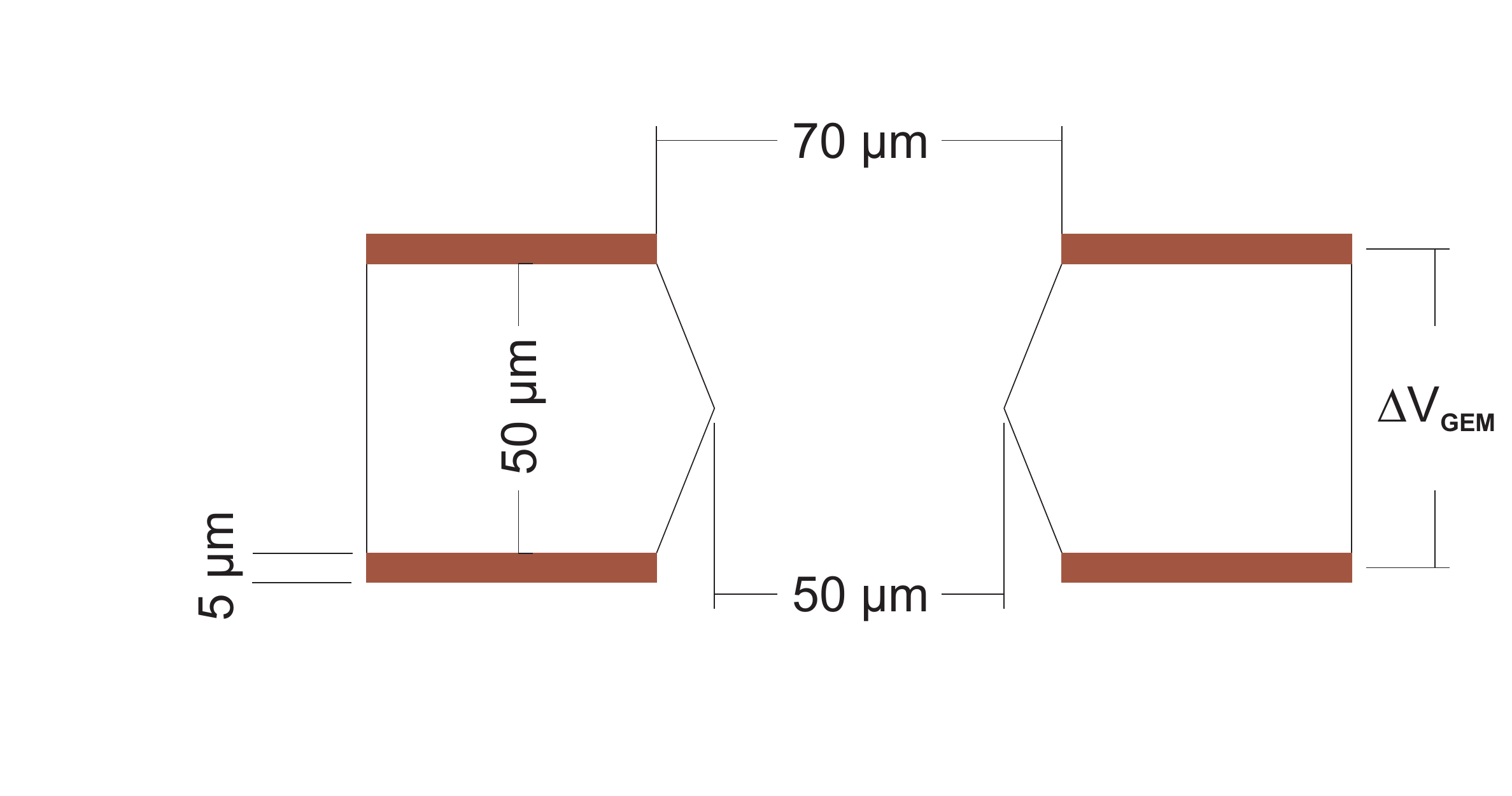}
\label{fig:cross_section}}\quad
\subfloat[]{\includegraphics[width=.45\textwidth]{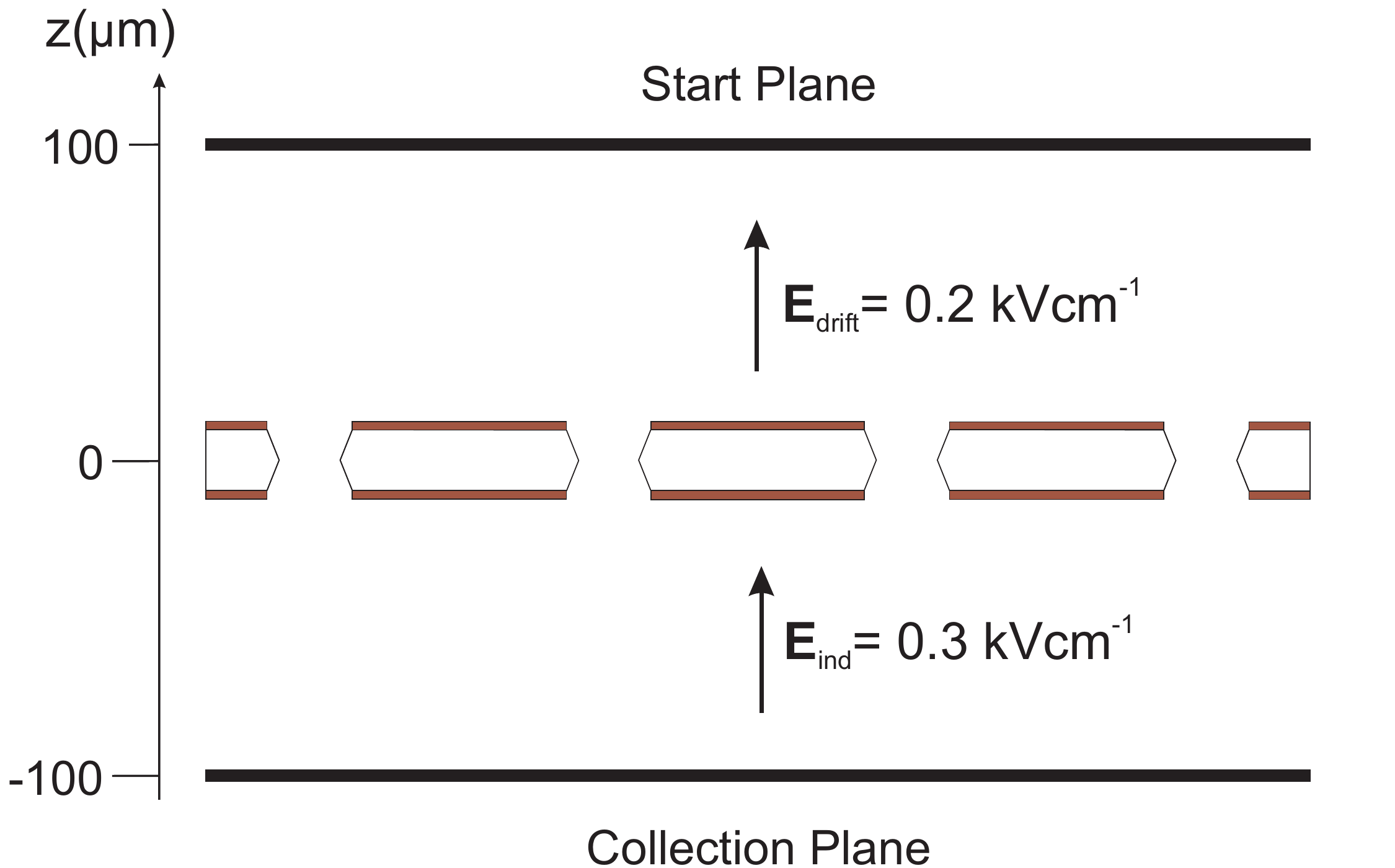}
\label{fig:plane_section}}
\caption{a) Cross sections of a GEM, with the dimensions and voltages between electrodes. b) Simulated configuration and applied drift and induction fields.}
\end{figure}

An hexagonal hole pattern with a distance between holes of 140 $\mu$m was considered, the insulator thickness is 50 $\mu$m and the metal electrodes are 5 $\mu$m  thick. Figure \ref{fig:cross_section} shows the GEM geometry.
The corresponding electric field configuration is depicted in figure \ref{fig:plane_section}. The hole has a bi-conical shape, the outer diameter is 70 $\mu$m, while the narrower part has 50 $\mu$m of diameter.

\subsection*{Gases}\label{subsec:gases}

In order to compare simulation results with our measurements, we simulated a gas mixture of $\mathrm{Ar \ 70\% \ / \ CO_2 \ 30\%}$. This is a penning-mixture, due to the presence of the quencher molecule CO$_{2}$, that opens new de-excitation ways for the previously excited noble gas molecules. If this excess of energy is above the ionization threshold of the quencher molecule, an ionization may occur, whit a probability called Penning probability. We used a Penning probability of 0.7 in this simulations based on previous calculations for $\mathrm{Ar \ 70\% \ / \ CO_2 \ 30\%}$ mixture\cite{Ozkan2010JINST,sven,penning1,penning2}.

The drift and induction fields (electric fields applied up and above the GEM, respectively) were 0.2 and 0.3 kVcm$^{-1}$.
Taking into account that the computational time strongly depends on the gain, because a higher number of electrons need to be tracked within the avalanche, a potential of 400 $V$ between electrodes was used for the firsts simulations tests, corresponding to gains of $\sim$10$^{2}$.
All simulations were performed considering a temperature of 293 K and a pressure of 760 Torr.

\section*{Simulation details}

\subsection*{Software platforms}

The Monte Carlo calculations involved three programs. Due to the complex shape of the GEM structure, an analytic solution for the electric field in the interest region is not possible to obtain.
To overcome this problem, the electric field is computed with \textit{Finit Element Methods} (FEM) software, that is used to calculate the electric potential in discrete nodes of a mesh, using boundary conditions. 
We used ANSYS$^\circledR$\footnote{www.ansys.com} to produce potential maps, to which we call generally as field maps, selecting the curved tetrahedral elements as our mesh elements, because they easily fit in sharp curved surfaces present in GEMs.

To simulate the drift and transport properties of electrons and ions in the MPGD gas medium, we used Garfield++ \cite{garfieldpp}. As input, this software requires the electric field configuration in the MPGD, the gas mixture, temperature, pressure and initial conditions of the primary charges (position, velocity and energy). 

In what regards the electric field configuration, we used Garfield++ to read the potential maps calculated with ANSYS$^\circledR$ and to calculate the electric field in any point of the space by interpolation between nodes.

A microscopic approach is used to simulate the drift of the charges. This uses Monte-Carlo methods to calculate the probability to occur each type of collision during the drift (elastic, excitation or ionization). 
The cross sections associated with each collision type are obtained from Magboltz\cite{magboltz,Biagi1999234}.

Primary electrons starts with assigned $\vec{r}_\mathrm{start}=(x_\mathrm{s},y_\mathrm{s},z_\mathrm{s})$, velocity $\vec{v}_\mathrm{start}=(v_\mathrm{x,s},v_\mathrm{y,s},v_\mathrm{z,s})$ and kinetic energy $E_\mathrm{start}$, drifting through the gas and producing secondary charges as it passes the multiplication region.
The final position of each secondary charge, $\vec{r}_\mathrm{end}$ and the effective gain are the observables of interested that are recorded for further analysis of the charging-up effect.

%

\subsection*{Initial attempts}
 
To start our simulations, we randomly distributed 10$^{4}$ primary electrons in the surface of a plane parallel to the GEM, located 100 $\mu$m above the GEM, indicated as the start plane in figure \ref{fig:cross_section}.

In order to determine the number of collected and deposited electrons and ions, the final position of each electron and ion from avalanches are analysed:
\begin{itemize}
  \item Electrons are collected if the final z is -100 $\mu$m below the GEM (the collection plane represented in figure \ref{fig:plane_section}.
  \item Ions are collected if the final z coordinate is in the top (positive) electrode of the GEM.
  \item Electrons and ions are deposited in the insulator surface if after the drift the z coordinate is between -25 $\mu$m to  +25 $\mu$m.
\end{itemize}  

\begin{figure}[htp]
  \centering
  \subfloat[Uncharged GEM]{\includegraphics[width=.45\textwidth]{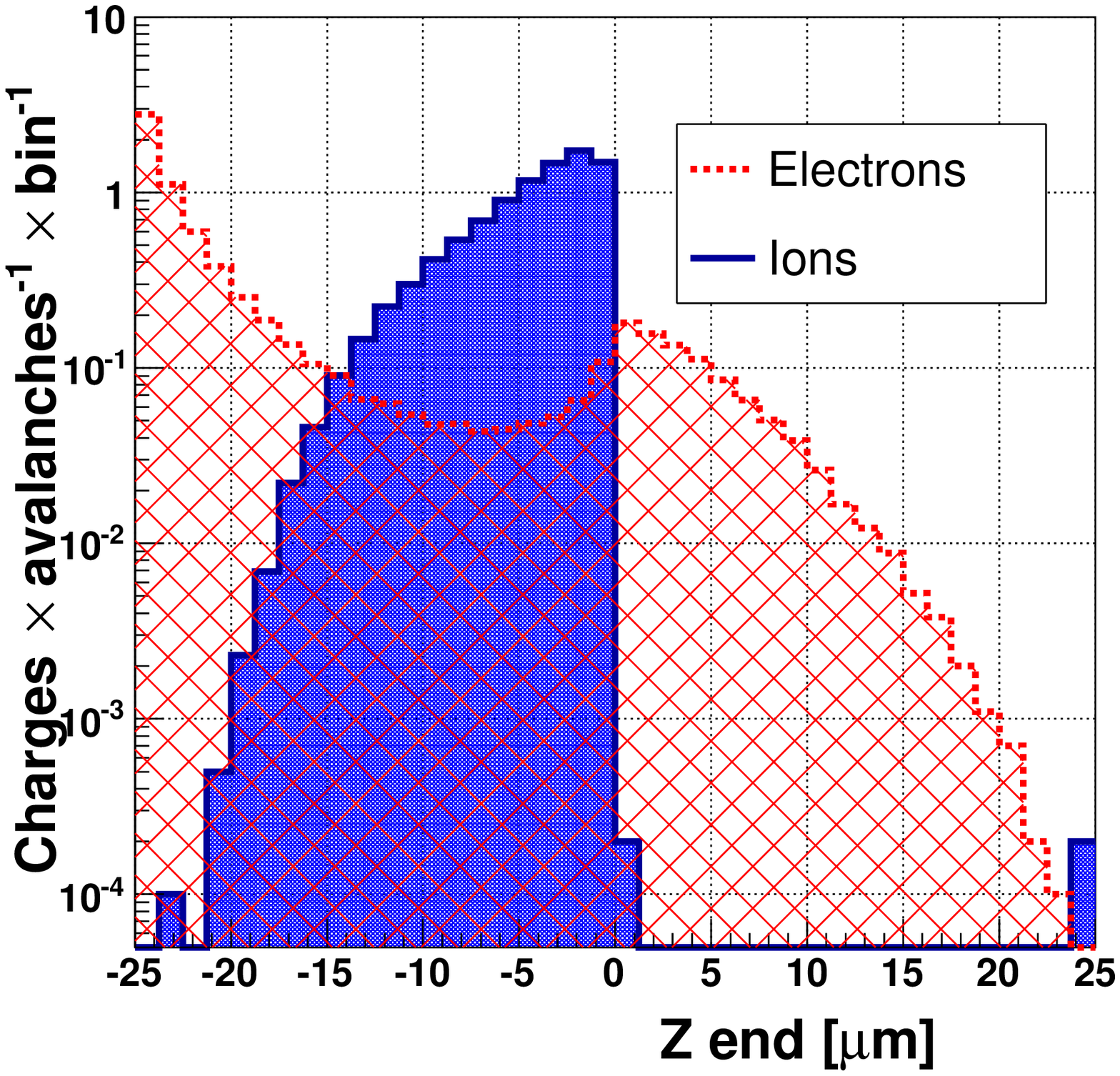}\label{fig:depo_hist_with_no_charg}}\quad
  \subfloat[Charged GEM]{\includegraphics[width=.45\textwidth]{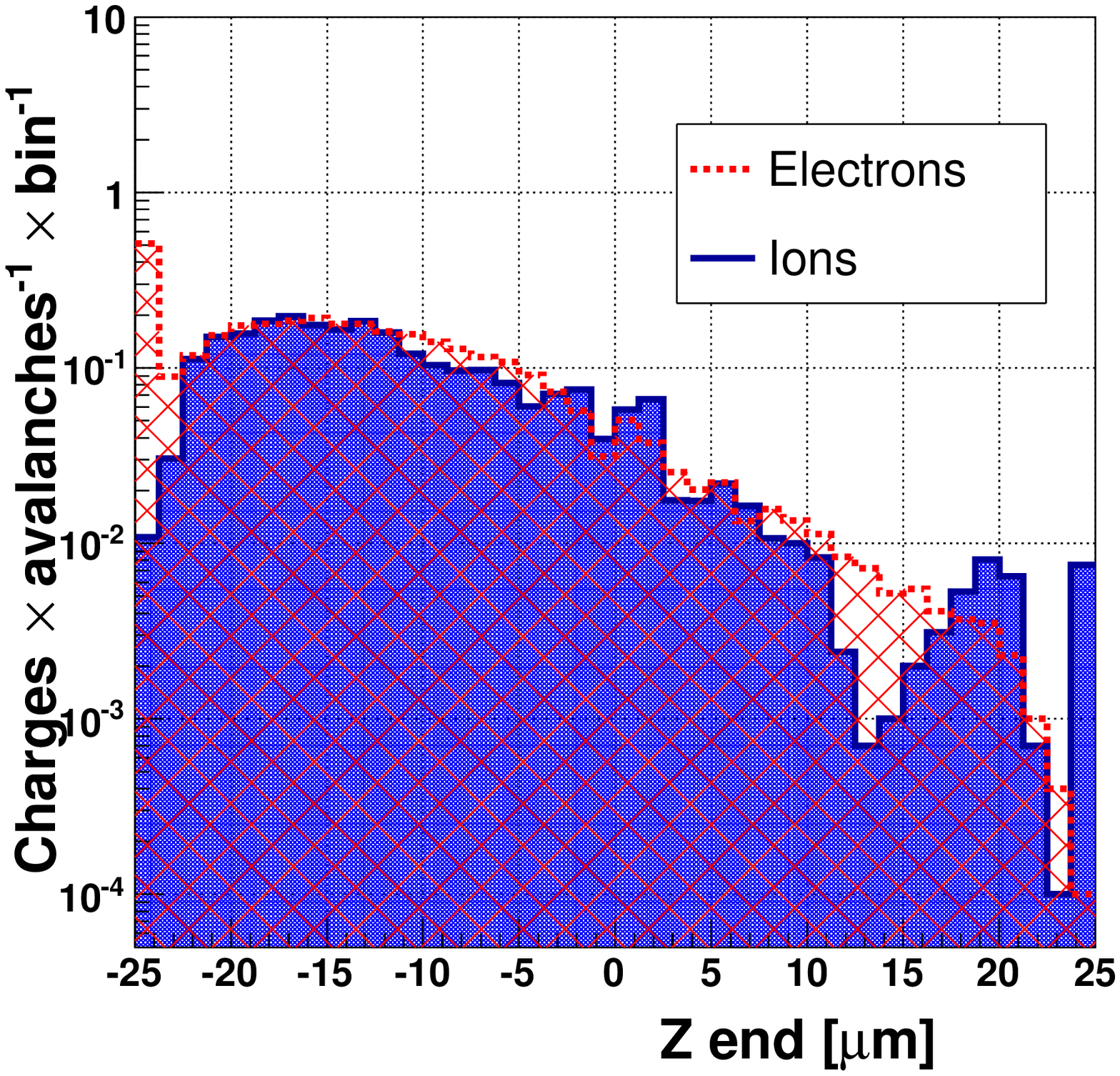}\label{fig:depo_hist_with_charg}}\\
  \caption{Spatial distribution of charges deposited in the insulator surface of the GEM detector, before (\ref{fig:depo_hist_with_no_charg}) and after (\ref{fig:depo_hist_with_charg}) simulation of primary avalanches, at $V_\mathrm{GEM}=$ 400 V.}
  \label{depo_hist_with_charg}
\end{figure}

GEMs previous to charging-up, i.e. without deposited charges, are called uncharged GEMs, being the charged GEMs those who have already deposited charges due to charging-up. 
The deposition distributions of charges (electrons and ions separately) in the insulator, for the case previous to charging-up (figure \ref{fig:depo_hist_with_no_charg}) shows that the charges are not deposited uniformly on the hole surface. 
In addition, the number of deposited electrons is higher than the number of deposited ions. The reason is related with the mass of each particle. Ions are heavier than electrons, and tend to follow very well the field lines, in the direction of the electrodes. Electrons has a much more chaotic movement, due to the lower mass, having more probability of ending in the insulator surfaces. This originates variations on the local electric field.

\
After some avalanches, the distribution of new electrons and ions that reaches the insulator tend to compensate each other, due to Coulomb attraction between previous and future deposited charges (figure \ref{fig:depo_hist_with_charg}). The local variation in the electric field will therefore vanish and a stable configuration will be achieved.

\begin{figure}[tbp]
\centering
\includegraphics[width=.3\textwidth]{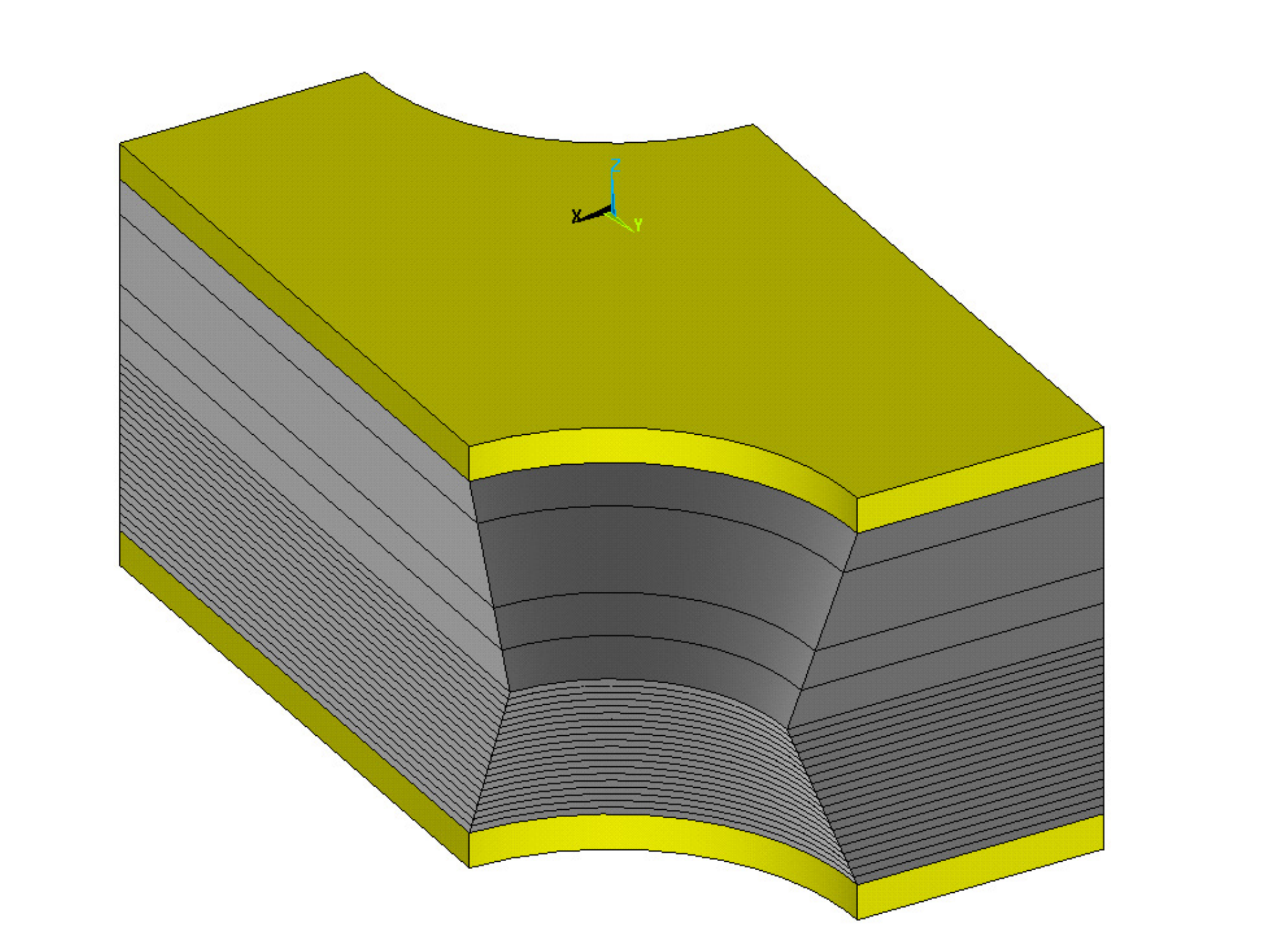}
\caption{Unity cell of a GEM, used to calculate the field maps with ANSYS$^\circledR$. 24 slices of different sizes were non regularly distributed due to the charge deposition non uniformity in the insulator surface, shown in figure \ref{fig:depo_hist_with_no_charg}.\label{fig:gem_slice}}
\end{figure}

In order to simulate the effective gain variation as avalanches happen, we needed to iteratively include this charge deposition in the potential maps computed with ANSYS$^\circledR$. The software does not provide the option to put single charges in their exact deposition position in the insulator surface. In addiction, this scenario would lead to discontinuities and numerical issues.
Instead, we created small slice surfaces in the insulator foil and add the correspondent density charge to each surface. Due to the shape of the deposition, and to computational limitations of field maps files for very small finite elements, we used 24 different slices in the insulator, achieving in this way a good balance between the detail of the calculations and the needed computing power. The slices are not regularly distributed, as shown in figure \ref{fig:gem_slice}, trying to match the z profile of the charge deposition histograms (figure \ref{fig:depo_hist_with_no_charg}).

\subsection*{Constant Step Method}\label{subsec:const_meth}

\begin{figure}[htp]
  \centering
  {\includegraphics[width=.5\textwidth]{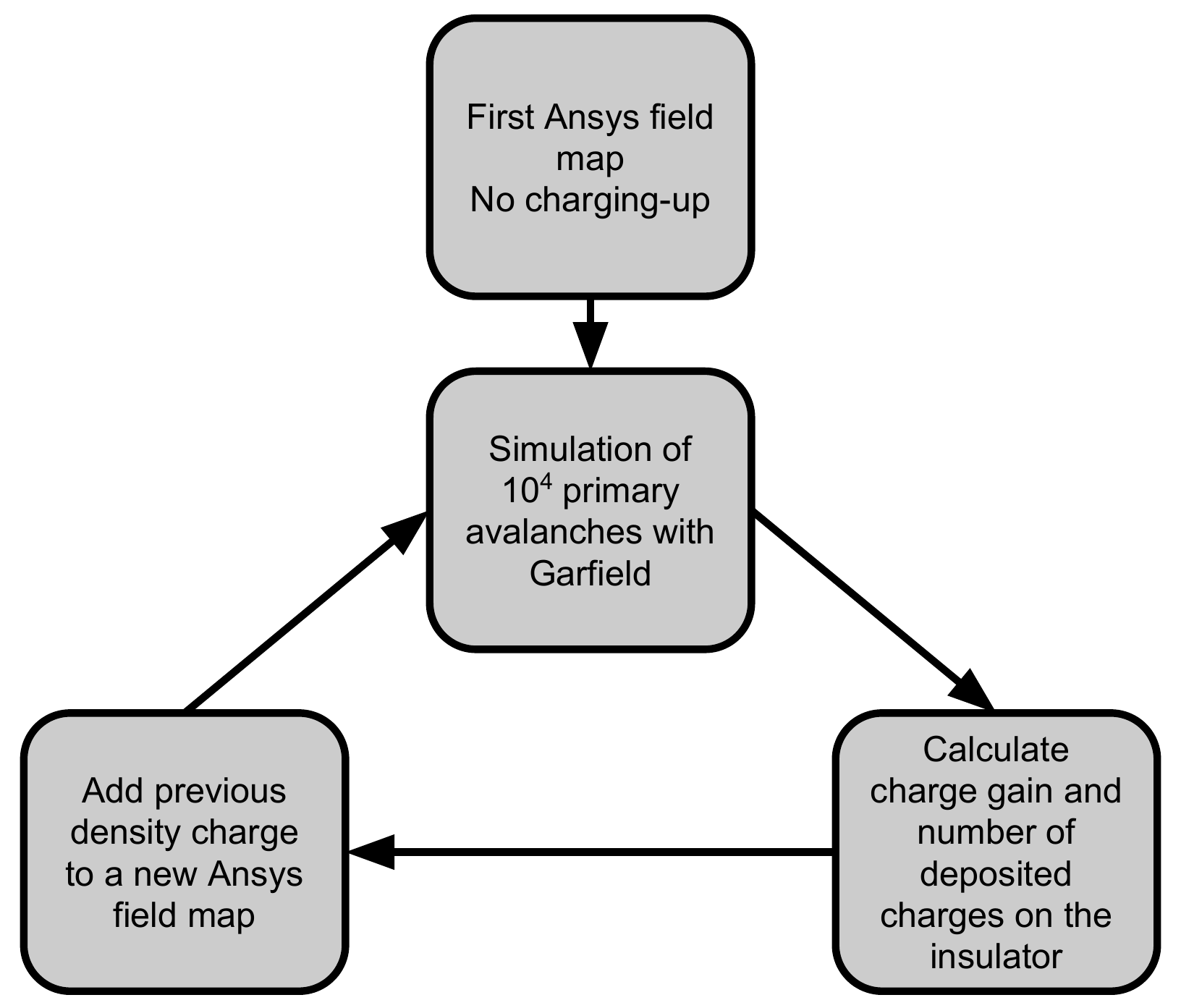}\caption{Constant step method flow-chart.\label{fig:cont_Step_meth}}}
\end{figure}

The flow-chart of the first iterative algorithm used to simulate charging-up iterations is in figure \ref{fig:cont_Step_meth}.

At the first iteration, we compute the electric field map assuming no charges in the insulator surface. Then, we import that field map into Garfield++, simulate $10^{4}$ primary avalanches and determine the density charge deposited in each insulator slice surface.
A new field map is created, with the contribution of previously deposited charges. The density charge in each slice is calculated taking into account the contribution of both the ions and the electrons ending up in the insulator surface. 
A new set of $10^{4}$ primary avalanches is simulated and the process is repeated iteratively.

It was found that statistical fluctuations in the calculated gain depends on the number of simulated avalanches per step, but the number of deposited charges per avalanche seems to be less sensitive to fluctuations.
A small step-size of $10^{4}$ primary electrons
was chosen in order to obtain good detail in the time evolution of charging-up. However, this small step implies hundreds of iterations until stabilization, which lead to a very heavy computation.

Since the number of deposited charges is the responsible for the local variation in the electric field, we use that observable as our control function for the iterative simulation, i.e. we stop our iterations when its value stabilizes over iterations (corresponding also to a gain stabilization).

\subsection*{Dynamic Step Method}

In order to accelerate the simulation process, we developed an extended method that uses a dynamic step-size in each iteration.
This step-size is smaller when the number of deposited charges per avalanche changes quickly, and is larger when this quantity is more constant, i.e. the deposition stabilizes.

\begin{figure}[htb]
  \centering
  {\includegraphics[width=.65\textwidth]{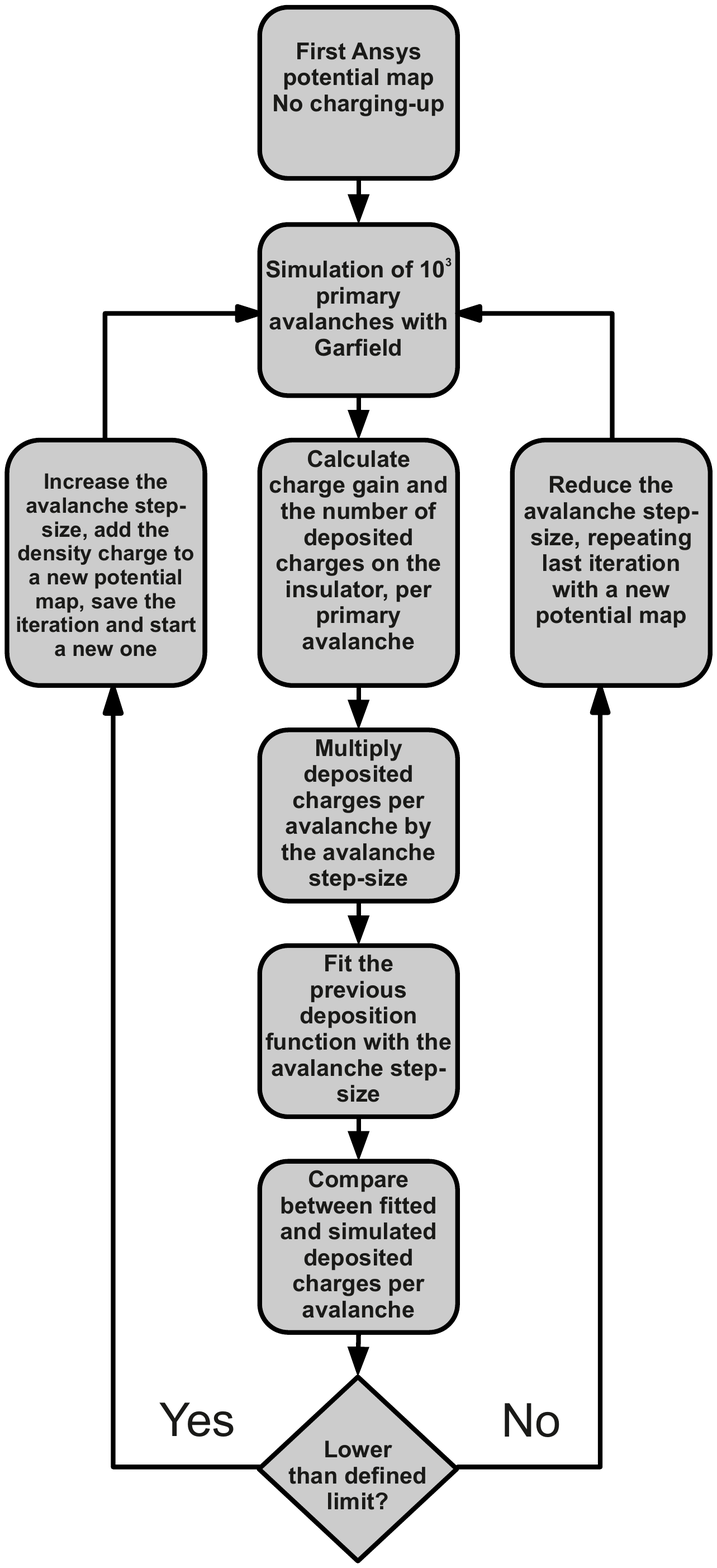}\caption{Dynamic step method diagram.\label{fig:din_Step_meth}}}
\end{figure}

To constrain the size of the step, we defined that the maximum total charge (sum of signs of ions and electrons) that can be added to the new field map should not be larger than 2$\times$10$^{4}$ $q_{e}$, (where $q_{e}$ is the elemental charge 1.6$\times10^{-19}$ C). This way, the maximum allowed number of avalanches per step is equal to $\frac{2\times10^{4}q_{e}}{G_{tot}}$, where $G_{tot}$ is the absolute gain in each iteration. The output of this calculation give us an upper limit for the step-size, considering a maximum charge that can be added to new potential maps in each iteration. 
Our attempts show that this upper limit value an acceptable value in terms of the convergence and speed of the method, but other limits can be defined.
The dynamic method is briefly described in the flow-chart in figure \ref{fig:din_Step_meth}.

The method starts with an uncharged ANSYS$^\circledR$ field map of the GEM. In each iteration we simulate $10^{3}$ primary avalanches, which is a good compromise between statistical fluctuations and computational time.

The number of deposited charges per avalanche, in each slice of the insulator surface, is multiplied by the variable step. 
For the firsts iterations, steps between $0.5\times10^{3}$ and $10^{3}$ primary avalanches were used. 

After the first 5 iterations (sufficient number of iterations that allows a reasonable fit), we fit the number of deposited charges per avalanche to a first order polynomial, and calculate for a given step, what the value of that function should be for the new iteration. 

We then simulate iteration number 6, and compare with the predicted value from the fit:

\begin{itemize}
\item If the difference between simulated and fitted value is larger than the maximum defined step, discard the iteration, the step is reduced to its half, and repeat the iteration.

\item If the difference between simulated and fitted value is smaller than the maximum defined step, the iteration is saved, we increase the step to the double. A new iteration is calculated and new fit considers only the last 5 valid iterations
\end{itemize} 

\section*{Results}\label{GEM_sub}
\subsection*{Comparison between methods}
\begin{figure}[htb]
\centering
\subfloat[]{\includegraphics[width=0.45\textwidth]{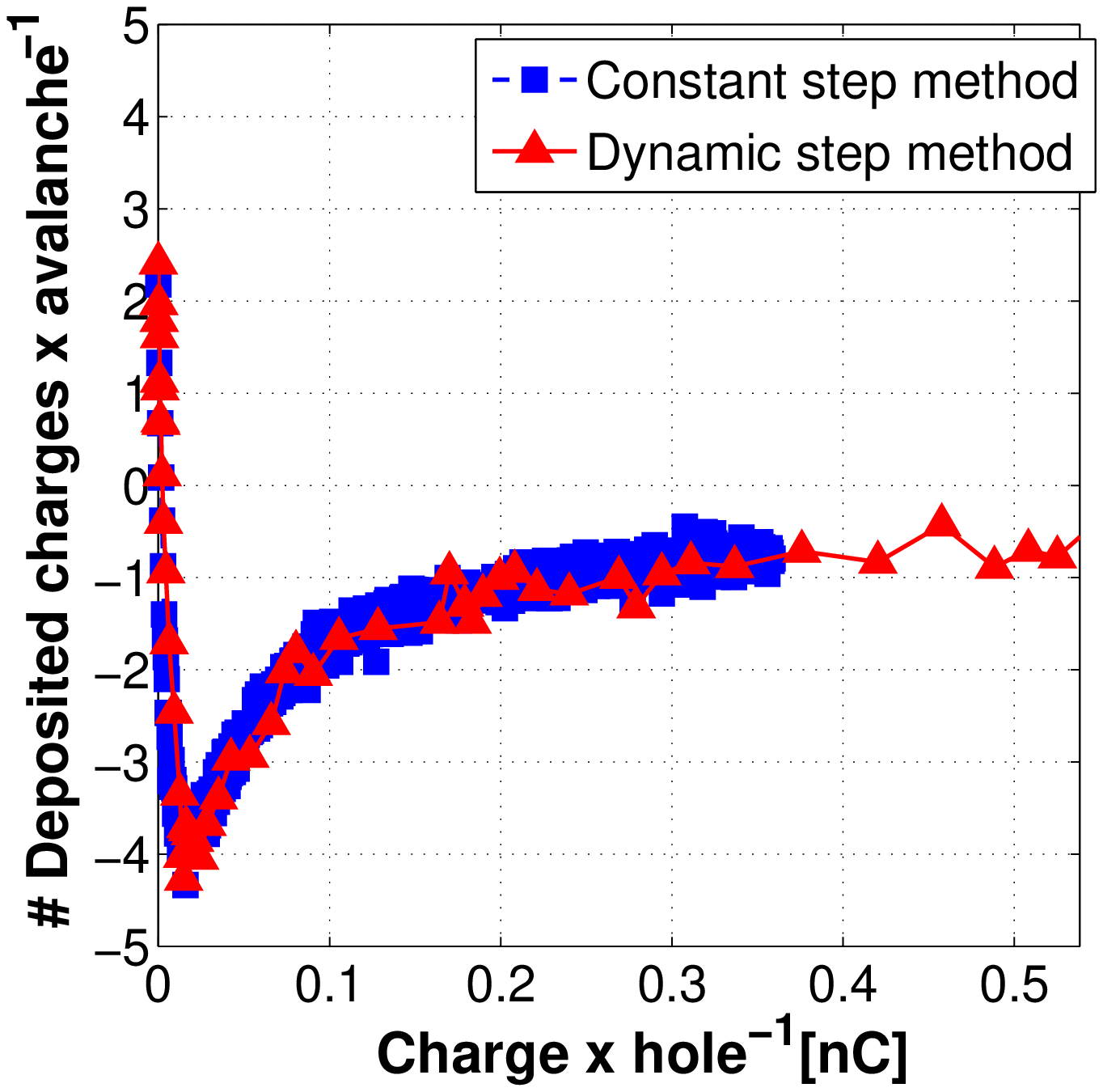}\label{fig:depo_2method}}\quad
\subfloat[]{\includegraphics[width=0.45\textwidth]{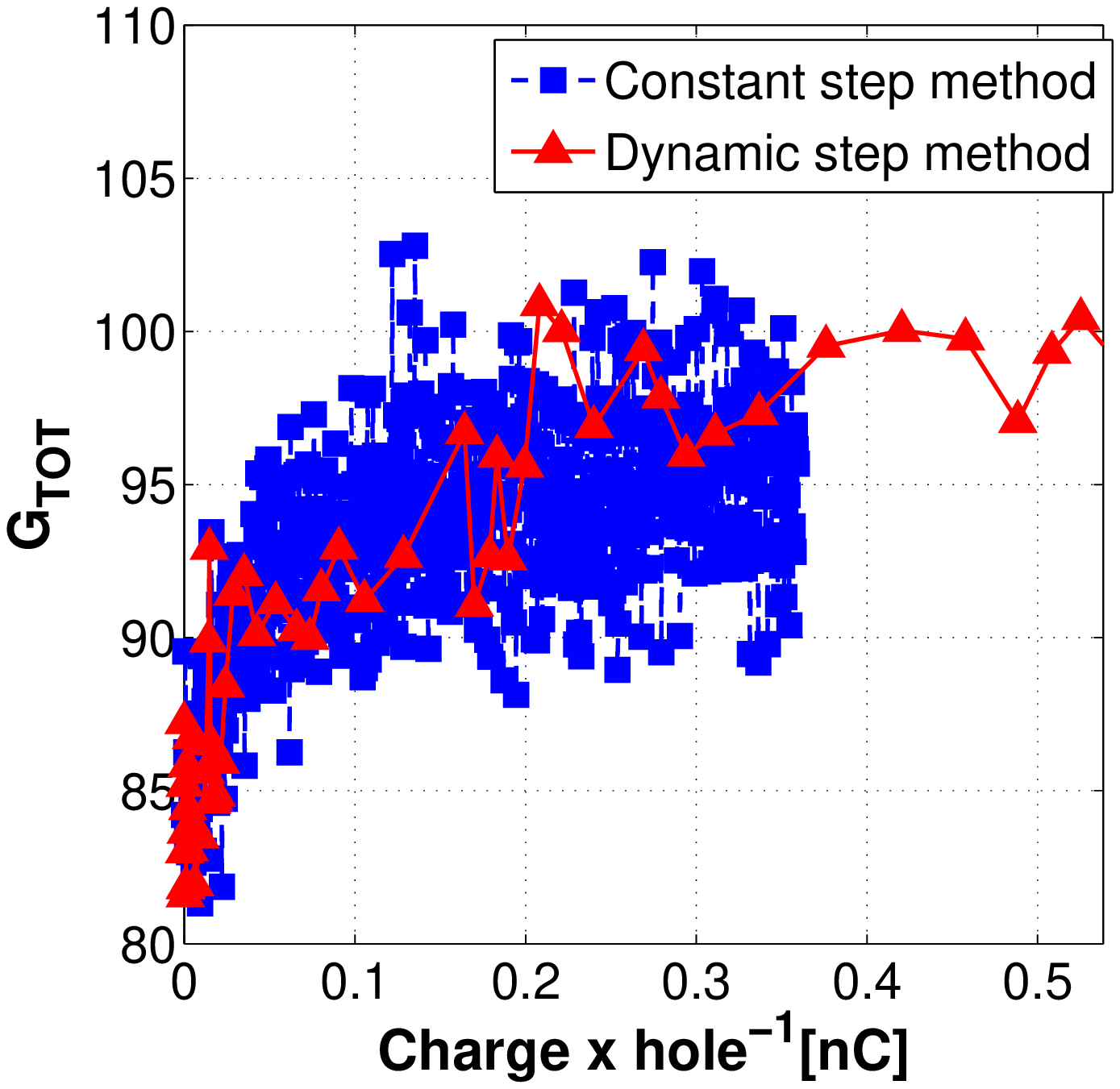}\label{fig:gain400V}}
\caption{a) Total number of deposited charges per produced secondary charge per hole, for both constant and dynamic methods. b) Comparison of the total gain (total number of secondary electrons produced per avalanche), between the constant and dynamic method. Both plots obtained for V$_{\mathrm{GEM}}=400$  V. \label{fig:compar}}
\end{figure}

The sum of all electric charges (the integral of the deposition histograms in figures \ref{fig:depo_hist_with_no_charg} and \ref{fig:depo_hist_with_charg}) deposited in the insulator surface, per primary avalanche, is shown in figure \ref{fig:depo_2method} for both methods, as function of the charge produced by each avalanche, per hole (is simply the number of primary simulated electrons in each hole multiplied by the total gain).

The agreement between both methods is clear. However, the dynamic-step method saves computational resources, using about one tenth of iterations.


Figure \ref{fig:gain400V} represents the total gain evolution for the two methods. We observe an increase in effective gain, followed by a stabilization plateau, reached in both methods. 
Due to previous results, from now on we will only consider the dynamic-step method for calculations.

\subsection*{Charging-up effect in the GEM transmission}
Primary electrons produced by incident radiation and drifting towards the GEM holes can be collected in the top electrodes, ending up not producing avalanches. 
The ratio between the number of primary electrons that enter the holes, producing avalanches and the total number of primary electrons simulated is defined as the electron transmission, shown in figure \ref{fig:trasmission}, for several voltages applied to the GEM electrodes.

The contribution of the charging-up effect in the electron transmission is more important when low voltages (<400 V) are used and negligible when higher electrical potentials are used.

\subsection*{Effective gain with and without charging-up}
\begin{figure}[htb]
\centering
\subfloat[]{\includegraphics[width=0.42\textwidth]{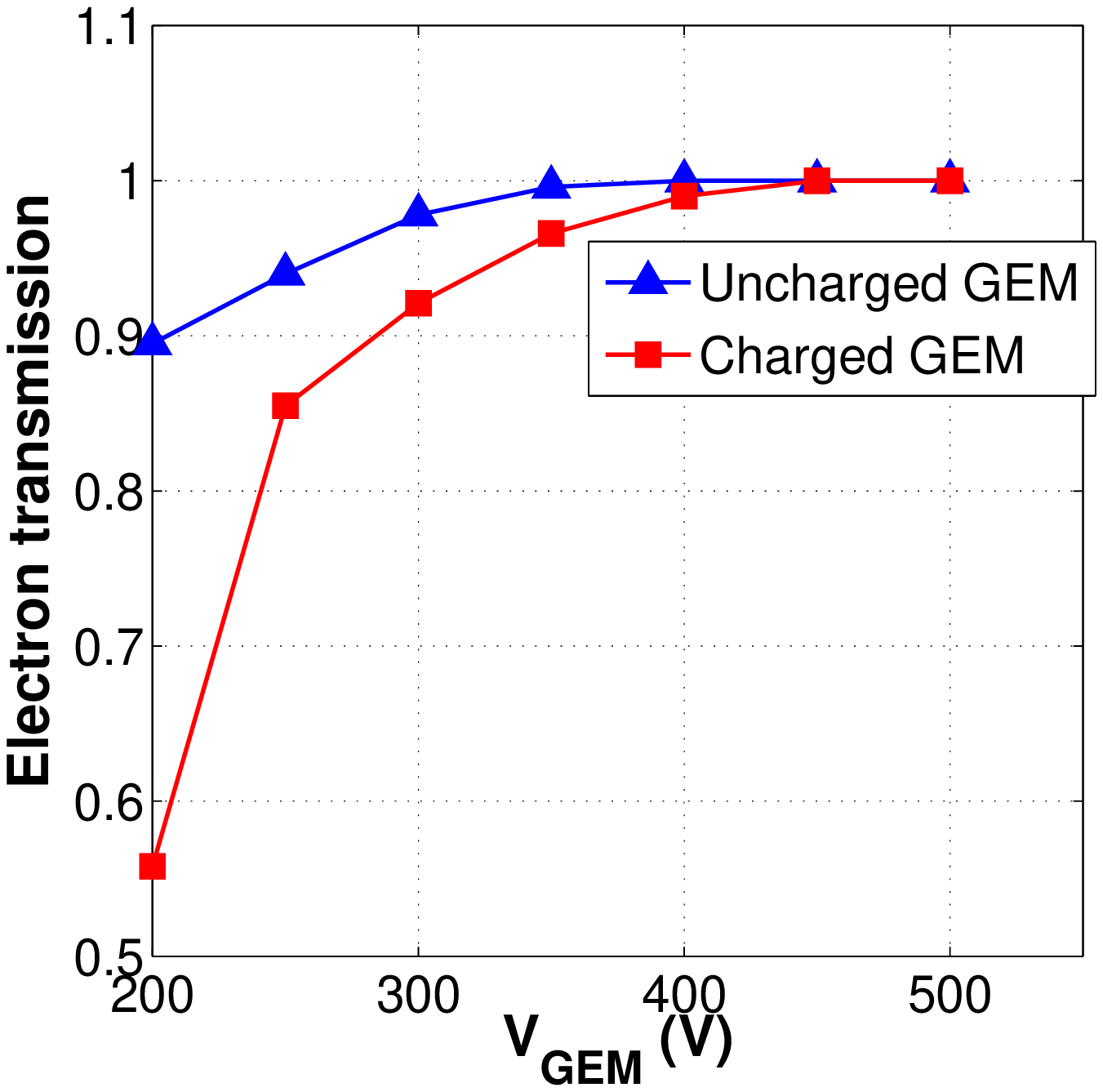}\label{fig:trasmission}}\quad
\subfloat[]{\includegraphics[width=0.43\textwidth]{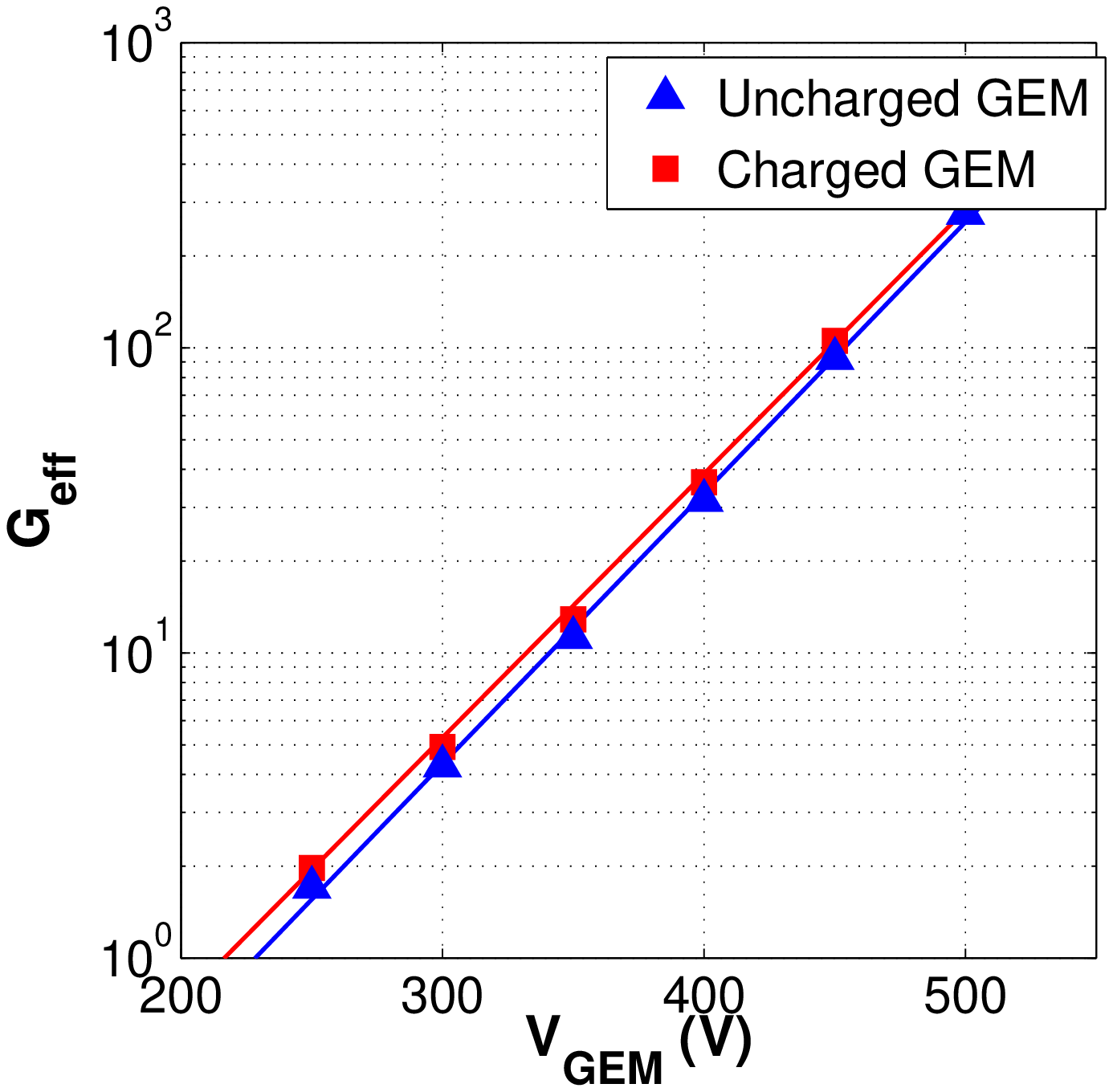}\label{fig:gain_vs_vgem}}
\caption{a) Electron transmission as a function of the voltage applied bettween the GEM electrodes. b) Effective gain comparison between charged (red) and uncharged (green) GEM, for different voltage between electrodes.}\label{fig:gain_and_transmission_with_charging}
\end{figure}

The dependence of effective gain on the voltage applied between electrodes in the GEM detector, is shown in figure \ref{fig:gain_vs_vgem}. 
The gain, after charging-up stabilization, is 10-15\% higher than the situation without charging-up.


\subsection*{Electric field intensity variation}
\begin{figure}[htp]
\centering
\subfloat[Without charging-up.]{\includegraphics[width=0.45\textwidth]{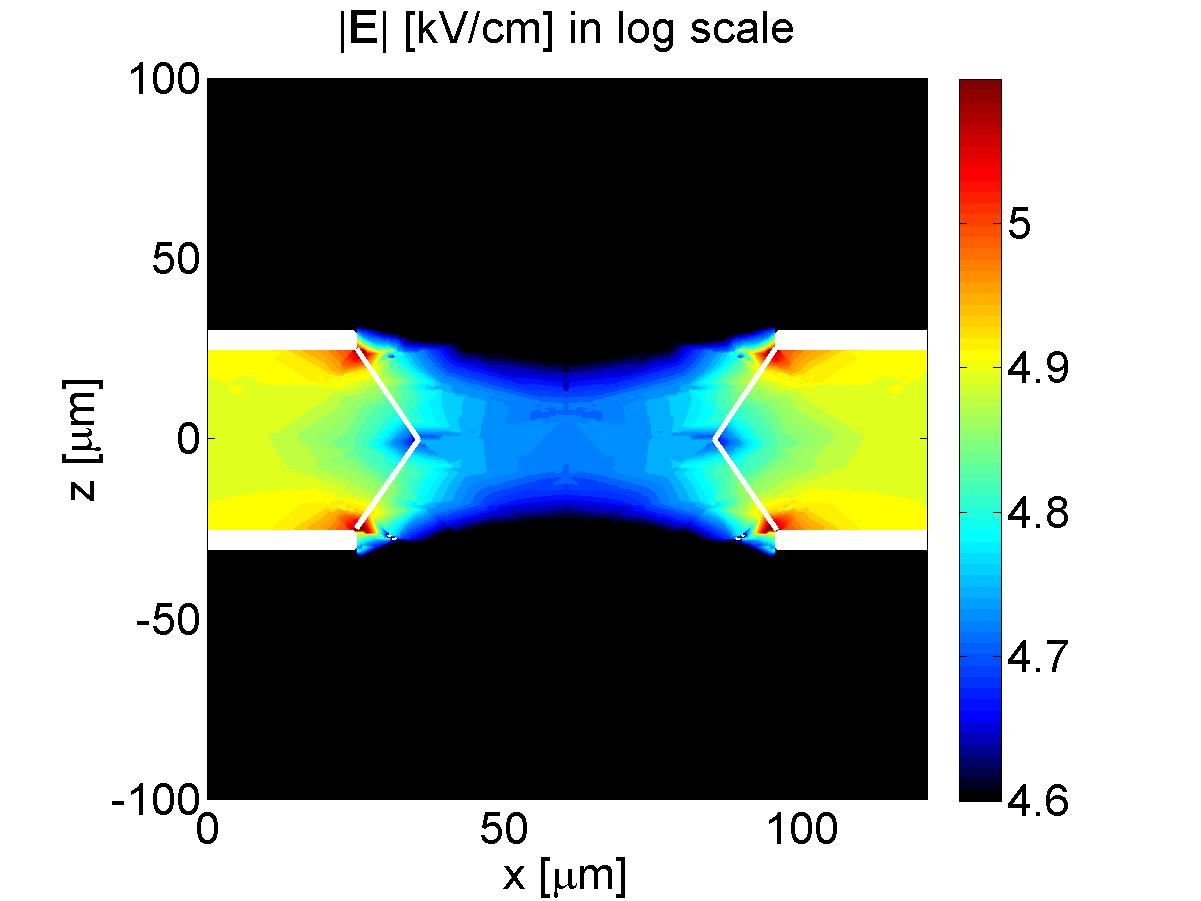}\label{fig:efield_a}}\quad
\subfloat[3 $\times 10^{6}$ avalanches.]{\includegraphics[width=0.45\textwidth]{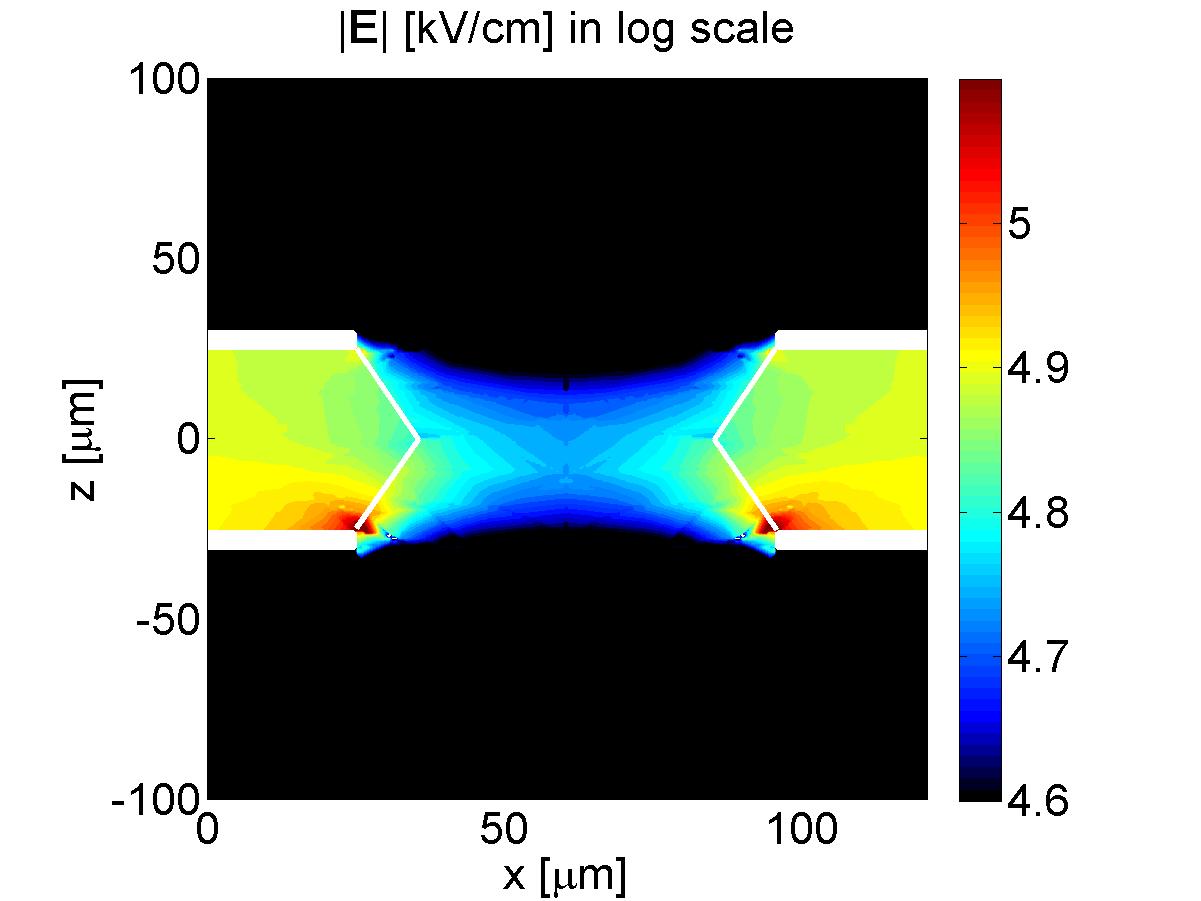}\label{fig:efield_b}}\quad
\subfloat[6 $\times 10^{6}$ avalanches.]{\includegraphics[width=0.45\textwidth]{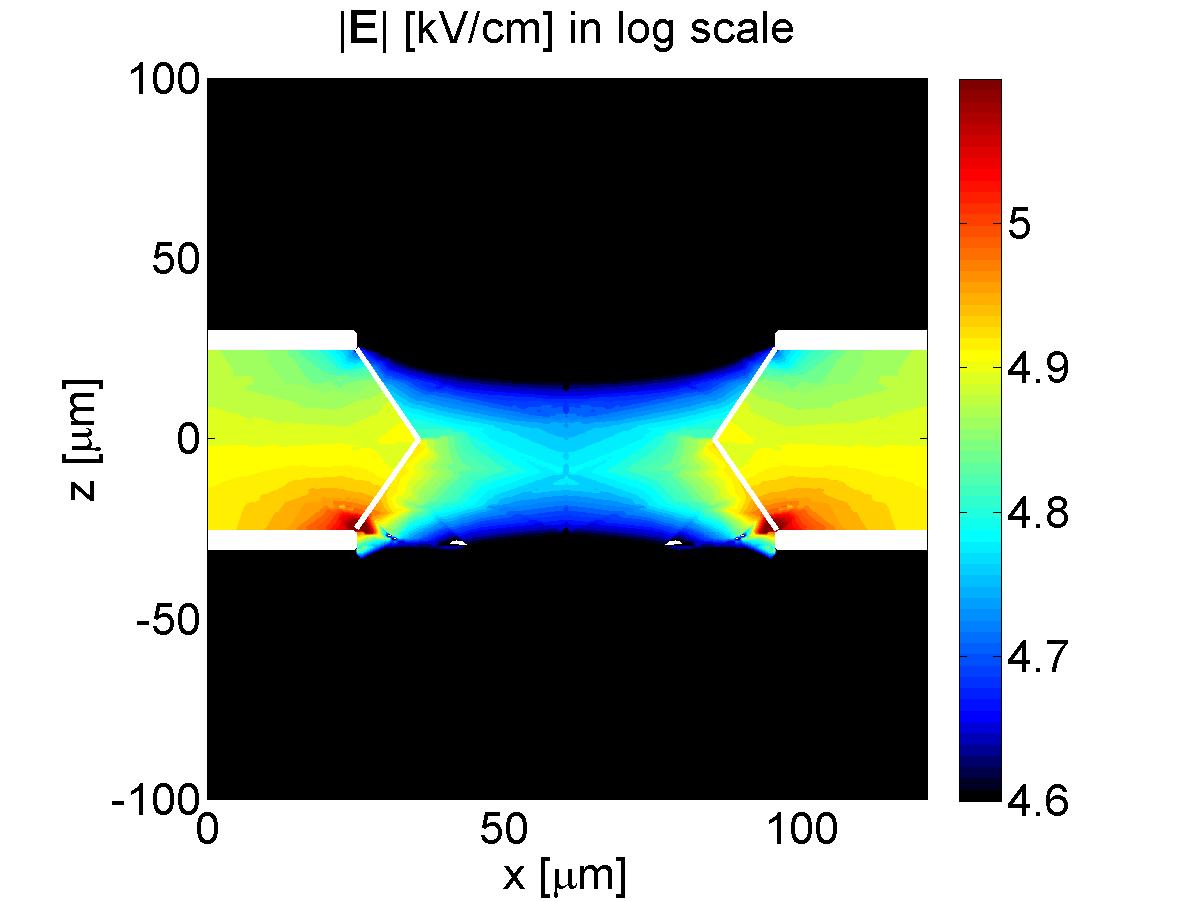}\label{fig:efield_c}}\quad
\subfloat[10 $\times 10^{6}$ avalanches.]{\includegraphics[width=0.45\textwidth]{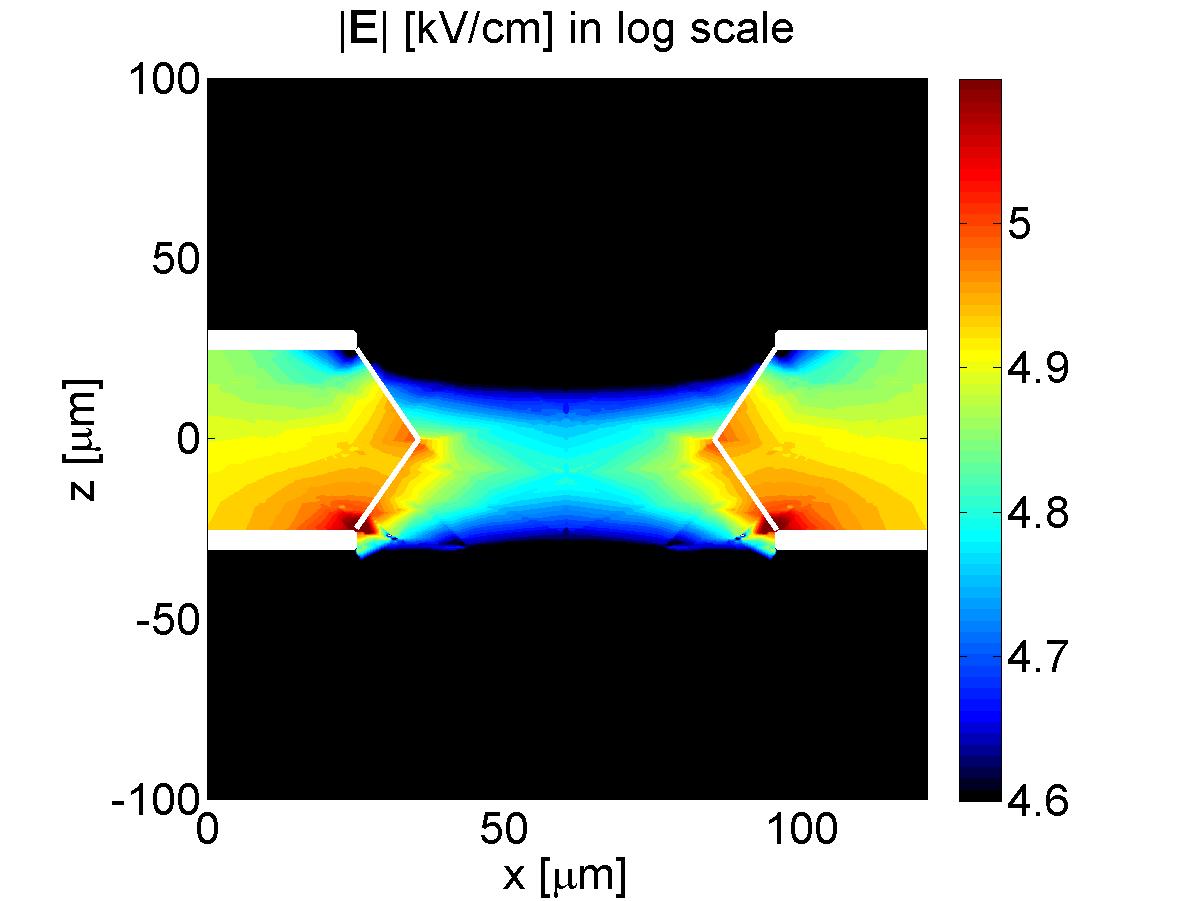}\label{fig:efield_d}}\quad
\caption{Evolution of the intensity of the electric field, in a GEM cross section. Computed with ANSYS$^\circledR$. The colorbar refers to the logarithm of \textbf{E}. Only intensities above 100 kVcm$^{-1}$ ($\ln (100)\simeq 4.6$) are colored. }\label{fig:efield_variation}
\end{figure}

A 2D representation of the electric field in the GEM is shown in figure~\ref{fig:efield_variation}. Each plot is obtained by calculation of the intensity of the electric field along a plane corresponding to a vertical cross section of the GEM hole, at four different stages of the charging-up process.

 We can observe that the biggest change in the electric field occurs near the electrodes. While the intensity of the electric field near the top (negative polarized) electrode decreases, it increases near the center of the hole and the bottom (positive polarized) electrode. 
The development of an avalanche inside the hole follows a nearly exponential model. The bigger fraction of secondary electrons is produced at the exit of the hole, in the last stages of the avalanches. There, the electric field is higher due to the charging-up effect, and thus, the effective gain increases as a result of this process.
 
 \subsection*{Comparison with experimental results}

Experimental measurements were performed at CERN. Physical parameters of the GEM, and the gas mixture used for measurements correspond to the simulation settings. X-ray photons were used as ionizing radiation. $\mathrm{K_{\alpha}}$ and $\mathrm{K_{\beta}}$ photons corresponding to energies 8.0 $\mathrm{keV}$ and 8.9 $\mathrm{keV}$ respectively were emitted by the X-ray tube which employed a copper target. 

\begin{figure}[ht]
{\centering \resizebox*{3in}{3in}{\includegraphics{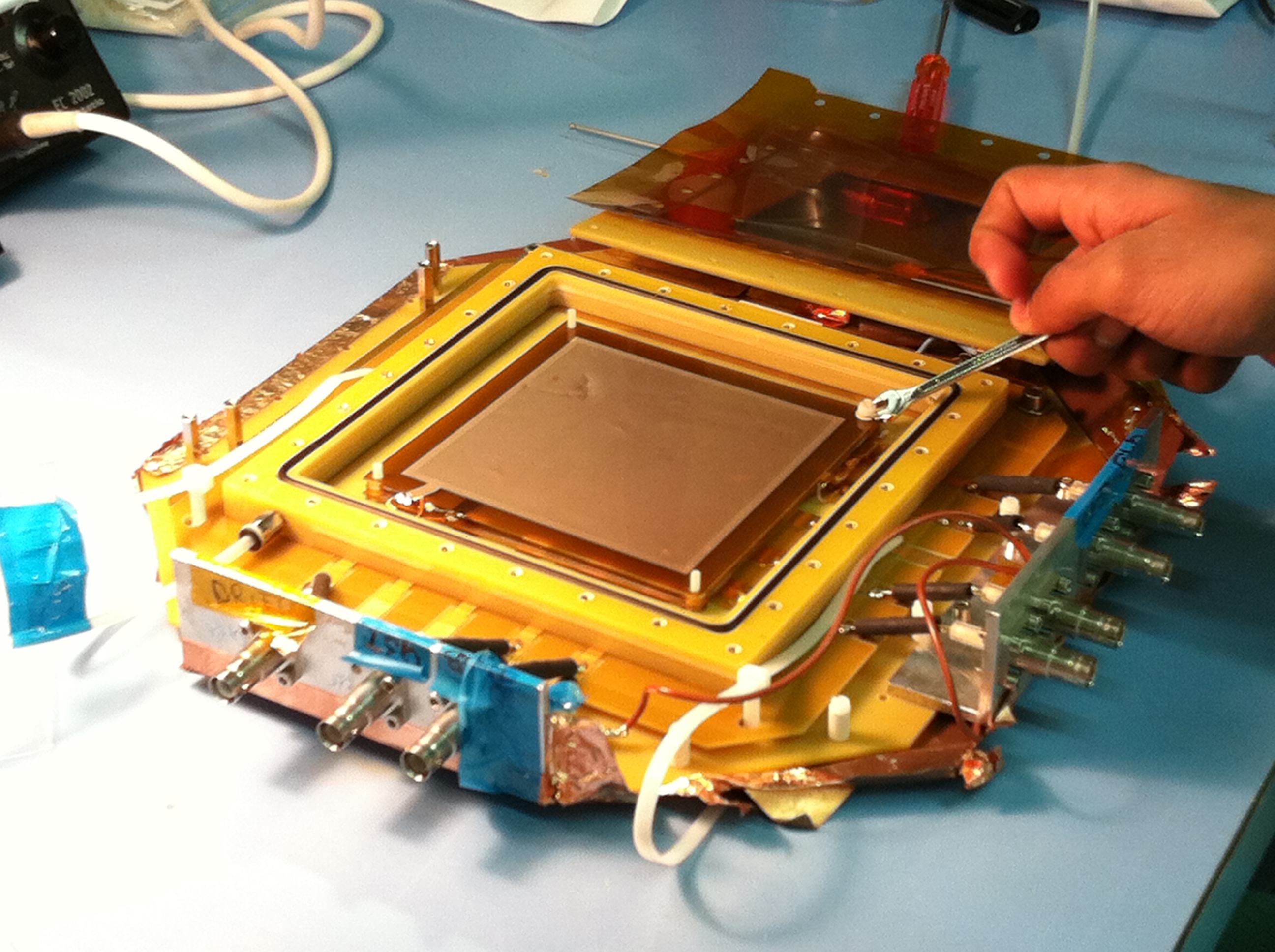}} \par}
\caption{GEM foil being mounted on a timing-GEM chamber}
\label{fig:Timing}
\end{figure}

\begin{figure}[ht]
\begin{minipage}[b]{0.5\linewidth}
{\centering \resizebox*{3in}{3in}{\includegraphics{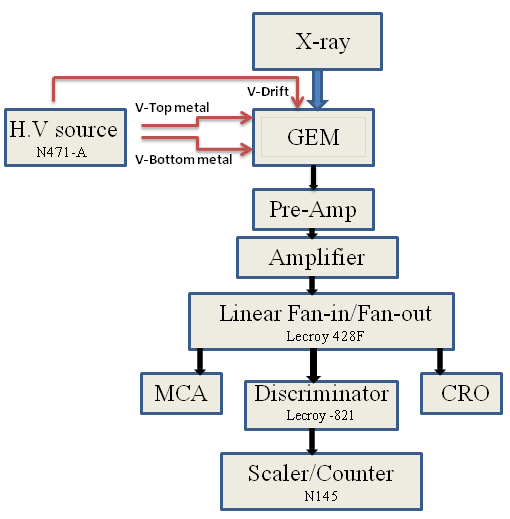}} \par}
\end{minipage}
\hspace{0.5cm}
\begin{minipage}[b]{0.5\linewidth}
{\centering \resizebox*{3in}{3in}{\includegraphics{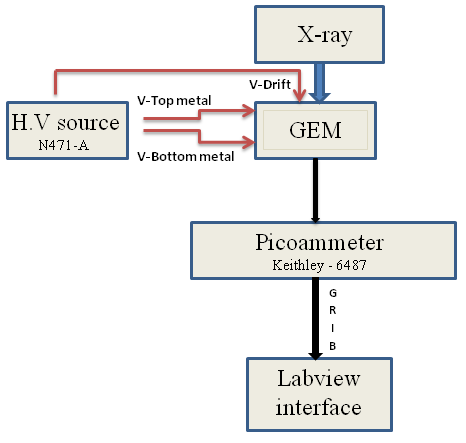}} \par}
\end{minipage}
\caption{Flow chart depicting the setup for Gain calibration (left) and Gain measurement (right).}
\label{fig:Flowchart}
\end{figure}

Collimators to control the photon flux were used in order to regulate the rate of charging up. The gain was measured over the time, for a constant irradiation flux. The GEM structure was housed inside an air-tight chamber, shown in figure \ref{fig:Timing}, that had a constant circulation of mixture, with a gas flow rate of $\mathrm{6\ l.h^{-1}}$. Chamber pressure was maintained at $\mathrm{760\ Torr}$. Figure \ref{fig:Flowchart} shows the schematics setup for gain calibration (left) and gain measurement (right). In order to compare with simulations, the experimental results were normalized from the time scale to charges per hole. This is done multiplying the number of primary electrons produced per second by the absolute gain, divided by the number of irradiated holes of the GEM.

For a detailed description of the measurement procedure refer to \cite{mythra}.

\begin{figure}[ht]
\centering
\subfloat[]{\includegraphics[width=0.42\textwidth]{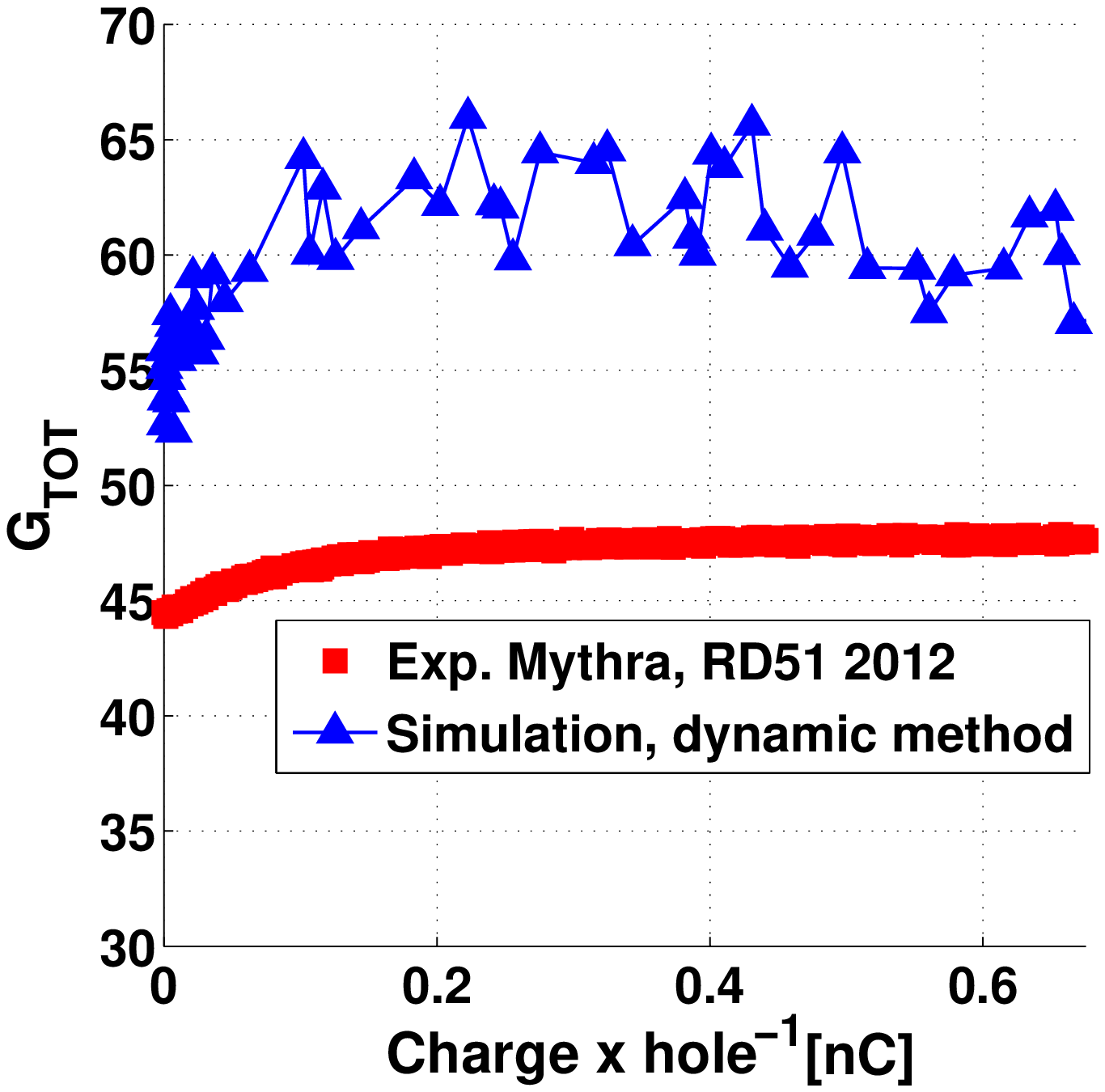}\label{fig:mythra_gain_comp}}\quad
\subfloat[]{\includegraphics[width=0.42\textwidth]{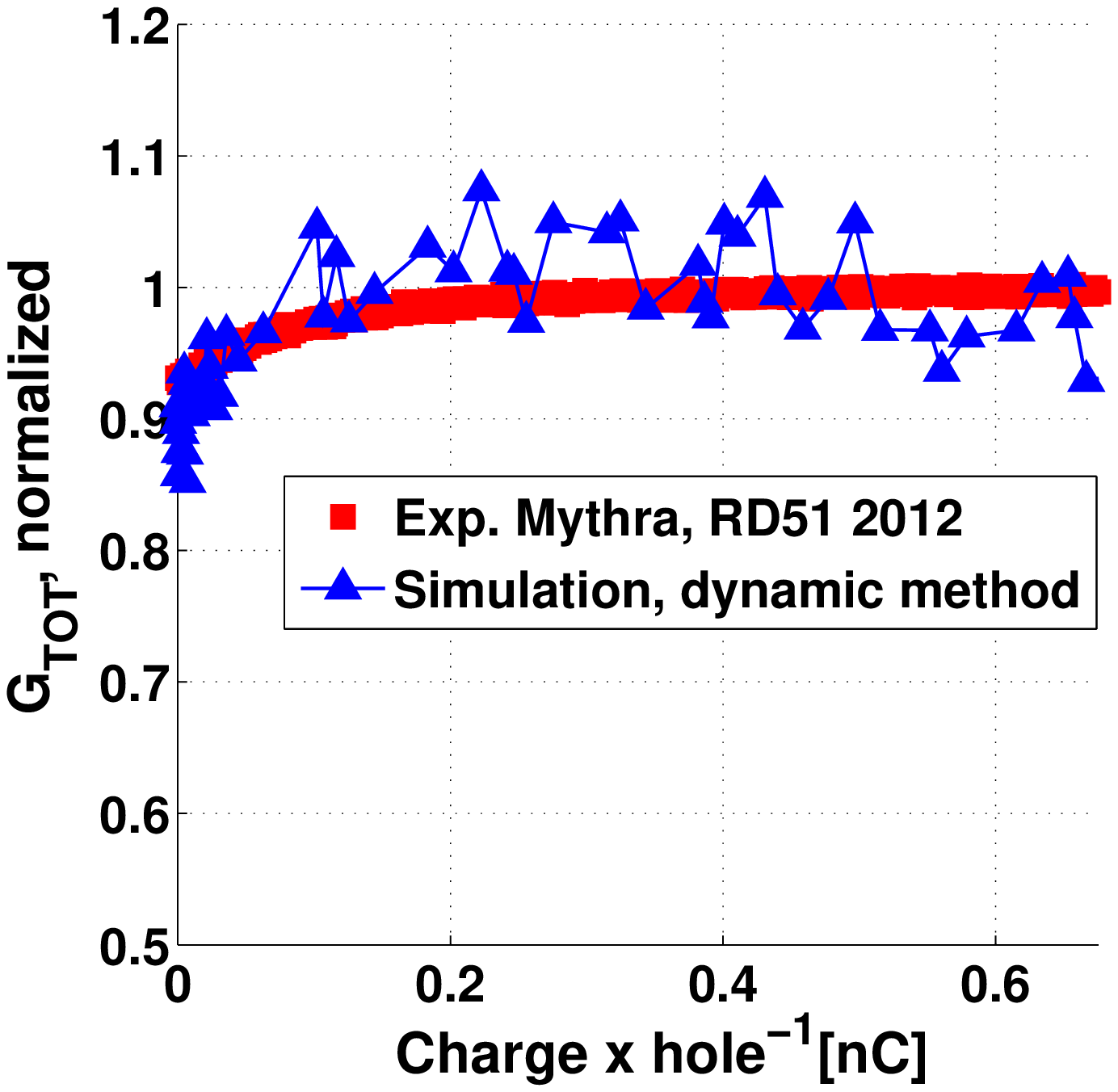}\label{fig:mythra_gain_comp_norm}}\quad
\caption{a) Absolute gain comparison between measuremments (red) and simulated (green) results. Same situation as figure \ref{fig:gain400V} but with $V_\mathrm{GEM}$=380 V. b) Same plot as figure \ref{fig:mythra_gain_comp} but with gain normalized. Experimental data taken by Mythra Varun Nemallapudi at RD51 facilities, CERN.\label{fig:mythra_gain_comp_tot}}
\end{figure}

A comparison between Monte Carlo calculations and measurements is shown in figure~\ref{fig:mythra_gain_comp}.
The total gain increases as the GEM starts to be irradiated, in both cases, achieving then a plateau.
A normalization to the plateau of both data was done to directly compare the time evolution of Monte Carlo simulations and experimental values, as shown in figure \ref{fig:mythra_gain_comp_norm}. From this observation, Monte Carlo simulations reproduce the time evolution of the gain.

On the other hand, the value of the total gain still do not match. This can be related with the mobility of the charges in the insulator surfaces and bulk, numerical issues associated with the finite elements method, computed with ANSYS$^\circledR$, impurities in gas and imperfections in GEMs dimensions introduced during the production.

\section*{Conclusion and future work}
In this work we have presented two iterative methods for the simulation of the insulator surface charging-up in GEMs, allowing a better understanding of their response.

Both methods agree between each other. However, the dynamic-step method saves computational resources.
The Monte Carlo functional time behaviour of the gain as the GEM is irradiated reproduces that observed experimentally.  
However, the absolute scale still not agree. 

Primary electrons transmission should be affected by the charging-up at lower voltages between electrodes of the GEM, but for higher voltages, used in regular applications, it does not play an important role.

Future work will include the application of the presented charging-up simulation methods to other MPGDs (e.g. THGEM) and the study of new geometries and detectors that could take advantages of this effect or minimize it.

The simulation of the mobility of deposited charges in the insulator surfaces could contribute to obtain more precise values. Refine the calculations of the electric field calculations (with ANSYS$^\circledR$ or other method) can also be important to get agreement between absolute simulated and measured gain values.

\acknowledgments
\noindent
This work was partially supported by projects CERN/FP/123604/2011 and  PTDC/FIS/110925/2009 through COMPETE, FEDER and FCT (Lisbon) programs.

\noindent P.M.M. Correia was supported by FCT (Lisbon) grant BIC/UI96/5496/2011.

\noindent C.D.R. Azevedo was supported by FCT (Lisbon) grant SFHR/BPD/79163/2011.

\bibliographystyle{PedroJINST}
\bibliography{charging-up-submission}

\end{document}